\documentclass[conference]{IEEEtran}
\IEEEoverridecommandlockouts
\usepackage[english]{babel}
\usepackage{booktabs}
\usepackage{cite}
\usepackage{amsmath,amssymb,amsfonts}
\usepackage{amsthm}
\usepackage{algorithmic}
\usepackage{graphicx}
\usepackage{textcomp}
\theoremstyle{definition}
\newtheorem{definition}{Definition}[section]
\theoremstyle{remark}

\usepackage{diagbox}
\usepackage{color}
\usepackage{xcolor}
\usepackage{colortbl}
\usepackage[flushleft]{threeparttable}
\usepackage{amsmath,amsfonts,amssymb}
\usepackage{tikz}
\usepackage{xspace}
\usepackage{bbding}
\usepackage{romannum}
\usepackage{paralist}
\usepackage{caption}
\usepackage{subcaption}
\usepackage[ruled,linesnumbered,boxed,vlined]{algorithm2e}
\usepackage{hyperref}
\hypersetup{
    colorlinks=true,
    linkcolor=blue,
    filecolor=magenta,      
    urlcolor=cyan,
    pdftitle={Overleaf Example},
    pdfpagemode=FullScreen,
    }
\usepackage{float}

\def\BibTeX{{\rm B\kern-.05em{\sc i\kern-.025em b}\kern-.08em
    T\kern-.1667em\lower.7ex\hbox{E}\kern-.125emX}}

\usepackage{amsmath}
\usepackage{amsfonts}
\usepackage{amssymb}
\usepackage{graphicx}
\usepackage{color}
\usepackage{xcolor}
\usepackage{enumitem}

\usepackage{url}
\usepackage[boxed,vlined]{algorithm2e}
\usepackage{listings} 
\usepackage{bbding} 
\usepackage[draft]{changes} 

\usepackage{tikz}
\usepackage{multirow}

\usepackage{diagbox} 

\newcommand{\hide}[1]{}
 \definecolor{brickred}{rgb}{0.8, 0.25, 0.33}

\newcommand{\bit}{\begin{compactitem}}
\newcommand{\eit}{\end{compactitem}}
\newcommand{\ben}{\begin{compactenum}}
\newcommand{\een}{\end{compactenum}}

\newcommand{\effective}{{Effectiveness\xspace}}
\newcommand{\automatic}{{Parameter-free\xspace}}

\date{}
\newif\ifblind

\theoremstyle{definition}

\begin{document}

\newcommand{\method}{\textsc{KeySelect}\xspace}
\title{Active Keyword Selection to Track Evolving Topics on Twitter}

\author{\IEEEauthorblockN{Sacha Lévy\IEEEauthorrefmark{1}\IEEEauthorrefmark{2},
Farimah Poursafaei\IEEEauthorrefmark{1}\IEEEauthorrefmark{2},
Kellin Pelrine\IEEEauthorrefmark{1}\IEEEauthorrefmark{2}, and
Reihaneh Rabbany\IEEEauthorrefmark{1}\IEEEauthorrefmark{2}
\IEEEauthorblockA{\IEEEauthorrefmark{1}School of Computer Science, McGill University}
\IEEEauthorblockA{\IEEEauthorrefmark{2}Mila - Quebec AI Institute}
}}
\maketitle

\IEEEpeerreviewmaketitle

\begin{abstract}
How can we study social interactions on evolving topics at a mass scale? Over the past decade, researchers from diverse fields such as economics, political science, and public health have often done this by querying Twitter’s public API endpoints with hand-picked topical keywords to search or stream discussions. However, despite the API's accessibility, it remains difficult to select and update keywords to collect high-quality data relevant to topics of interest. In this paper, we propose an active learning method for rapidly refining query keywords to increase both the yielded topic relevance and dataset size. We leverage a large open-source COVID-19 Twitter dataset to illustrate the applicability of our method in tracking Tweets around the key sub-topics of \textit{Vaccine}, \textit{Mask}, and \textit{Lockdown}. Our experiments show that our method achieves an average topic-related keyword recall 2x higher than baselines. We open-source our code along with a web interface for keyword selection to make data collection from Twitter more systematic for researchers.
\end{abstract}

\section{Introduction}
\label{sec:intro}
Twitter is a rich real-world data source, not only for its large user base and numerous conversations but also for its developer toolkit allowing programmatic retrieval of large quantities of Twitter data. Researchers have employed the Twitter API to study a wide variety of subjects, ranging from spontaneous events such as Hurricane alerts \cite{kejriwal2018pipeline}, to long-lasting global scale phenomenons, such as the COVID-19 pandemic \cite{banda2020largescale, alqurashi2020large, Chen_2020}, or the US 2020 Elections \cite{chen2020election2020}. Although these events yield datasets of radically different natures, through their sizes and time-spans, researchers leveraging them all rely on high-quality data collection processes. In order to accurately track their topic's evolution, and to maximize the amount of conversation retrieved, the topic relevance of each tweet matters. For example, when researchers are attempting to predict flu trends \cite{achrekar2011predicting}, missing a small bubble of individuals discussing the topic of interest may significantly impact the outcome of the experiments. Twitter offers public API endpoints allowing researchers to query the platform's databases using keywords, timestamps, or even geographical coordinates. The social network offers a free API tier, enabling researchers to conduct rate-limited queries every month, enabling access to a small number of past tweets through a search feature, and a real-time sample of the tweet flux through a stream feature \cite{jurgens2016tutorial}. Although practical to use, these API endpoints become quickly limiting when attempting to retrieve large amounts of past data, or actively tracking a niche discussion topic. The quality of the datasets yielded through the usage of these recovery methods are contingent on two primary factors: first, the underlying mechanisms of the Twitter API at work to generate both searches, and stream tweets samples, and secondly, the queries crafted by researchers to retrieve the content of interest to their topic. Although the black-box sampling operated by Twitter remains a concern for researchers \cite{pfeffer2018tampering, morstatter2016tamper}, the importance of the process of selecting optimal keywords for data collection queries seems to be under explored in the current literature \cite{bozarth2022keyword}.

Indeed, researchers who collect Twitter datasets focusing on particular topics have taken various roads while selecting keywords and query items \cite{alqurashi2020large, qazi2020geocov19}. A trivial way of selecting keywords is to perform a manual search on the platform's website: this process is quick and accessible but constrains the researcher to a very narrow perspective on their topic of interest (notably due to the personal curation of feeds by Twitter). Another, more complex way, relies on selecting expert users \cite{ghosh2013sampling} which can then be used as query items to retrieve connected, topic-relevant users (notably through follower-followee relationships), or to extract topic-relevant keywords by processing expert users' tweets. In either case, the limitations imposed by the Twitter API deprive researchers of a ground truth on their topic's discussion spaces (whether about active users, or related keywords). Users of the Twitter API are inherently limited by this partial point of view, and can at best circumvent the lack of perspective on the ongoing discussions with a thorough application of their domain knowledge during the query building process (while selecting keywords, and picking expert users). A convenient topic to study these phenomenons is the COVID-19 pandemic. For the past two years, COVID-19 has had a dominant role in Twitter conversations, impacting interactions regardless of the political spectrum, or nationalities \cite{Shuja2020.05.19.20107532}. Due to its breadth and depth, the topic of COVID-19 has yielded many sub-topics, arguably linked to the ramifications of the pandemic in social, economic, and political settings. Examples of these ramifications are the implementation of lock-downs across the world since early 2020, the deployment of mask mandates, and more recently, the distribution of vaccines. Discussions of these sub-topics have had complex timelines since the inception of the pandemic. In order to conduct accurate analysis using Twitter data, such as to measure the impact of the virus on political polarization \cite{yang2021online}, it is important for researchers to craft representative Twitter queries in order to capture a maximum of topic-relevant tweets.

 To the best of the authors' knowledge, there are only a few works studying the selection of topic-relevant keywords \cite{bozarth2022keyword}, notably through focusing on the application of text-based methods to tweet corpora \cite{marujo2015automatic, king2017computer}. In this paper, we first demonstrate the impact of query keywords onto retrieved Twitter datasets in Section \ref{background}. By exploring small samples of three independent large-scale datasets \cite{Chen_2020, banda2020largescale, qazi2020geocov19} we observe that retrieved tweet sets show significant differences, along with the keyword sets used to hydrate them. We then define the task of \textit{Active Keyword Selection} in Section \ref{sec:meth}, and present our method, \method, as a solution to the keyword selection problem. The selection process is structured in rounds and is iterative by default. On every iteration, an oracle is prompted for a positive or negative label on a candidate keyword, which may then be used to guide the keyword suggestion process. We formulate this issue as a matter of retrieving a maximum number of topic-relevant keywords. The manipulation of tweet sets required to construct data structures for the operation of \method and its baselines are also formulated in Section \ref{task}. We offer definitions of precision, recall, as well as volume and user coverage to evaluate the performance of a given method while actively selecting keywords. The performance of the presented method is illustrated through a comprehensive set of experiments leveraging 21 million tweets over the month of April 2020 from \cite{Chen_2020} by putting a focus on the three sub-topics of \textit{Vaccine}, \textit{Mask}, and \textit{Lockdown}. Using the defined \textit{Active Keyword Selection} task and measures, we compare \method against baselines for multiple oracle labeling budgets. Our method shows better performance in terms of recall and precision than baselines while yielding a higher volume of tweets.
We summarise our contributions as follows:
\begin{itemize}
    \item We illustrate the current challenges of keyword selection to collect high quality Twitter datasets, using the search and stream Twitter APIs. More specifically, we show that large scale datasets collected by independent teams to study COVID-19, do not overlap closely and capture different discussions, mainly due to their hand-crafted keywords. 
    \item We frame the task of automatic selection of keywords to follow and capture a topic of interest, and present an active learning method to help researchers rapidly select topic-relevant keywords using a set of tweets and their few initial best-guess keywords.
    \item We demonstrate that \method can yield up to a 25\% increase in retrieved volume of users and tweets on a given topic, with an average recall value 2.8 times greater than its baselines.
\end{itemize}
We open-source the code for our method and experiments at{
\color{red}\url{https://github.com/sachalevy/active-keyword-selection}}. We also provide a simple web interface for researchers to easily use our keyword selection method. All software is provided under the MIT license.

\section{Related Work}
\label{sec:relatedwork}
We consider related work relevant to the data structures employed in this paper (sampling on graphs), as well as the underlying tasks to be performed (query building and discovering topics).

\subsection{Graph Sampling}
Our method generates tweet-hashtag and user-hashtag networks from a set of tweets. These graph are then sampled to be analysed. There exist multiple methods based on Random Walk or Metropolis-Hastings algorithms which allow to efficiently sample large graphs. As demonstrated by \cite{leskovecfaloustos2006, hu2013survey}, using a Random Walk algorithm to sample nodes from large graphs is more efficient than Monte-Carlo-based alternatives in order to generate a sub-graph with similar features to its parent. \cite{wang2011understanding} applies these sampling algorithms to the case of large-scale social media graphs, and finds that no satisfying sampling can be achieved while preserving the properties of full-scale graphs. Moreover, they note that the performance of the Metropolis-Hastings random walk highly depends on the experimental dataset. Although very general, these sampling methods cannot be directly applied to generate sub-graphs with transformed characteristics (for example, by biasing towards hashtags focusing on a topic of interest). We compare our keyword selection method to the Random Walk algorithm to select candidate topic-relevant keywords to suggest to the oracle. 

\subsection{Query-based Search}
Filtering keyword selection for Twitter is related to a database query problem. The recent emergence of large databases, able to hold billions of records, induced the issue of efficiently retrieving content of interest. This problem is treated by \cite{stryker2006}, which defines a methodology to query a news article database for cancer-related records. The authors attempt to find a set of query terms that maximize retrieved articles mentioning cancer, and conversely, minimizing those omitting the subject. The authors define precision as the ratio of retrieved related articles to total retrieved articles. Furthermore, they define recall as the ratio of retrieved related articles to total available related articles. Here, the authors have direct access to the whole database, such that they are able to estimate these two measures for a given set of search terms. Working by iterations, they begin with a simple set of common words related to cancer. They manually label the recall and precision of various search terms, gradually improving the quality of their queries. The quality of this manual labeling is also tested using Krippendorff's Alpha-Reliability measure  \cite{Krippendorff2011ComputingKA}. We formulate the keyword selection for the Twitter data collection problem in similar terms as done by \cite{stryker2006} for electronic database query terms. Note that we do not have access to all Twitter data, which imposes restraints on our ability to measure recall on a set of selected keywords. In Section \ref{sec:dataset}, we argue that our choice of experimental dataset and study sub-topics approximates a complete Twitter universe.

\subsection{Topic Discovery}
Capturing the essence of a topic in real-world corpus of text is a challenging task. Our method strives to suggest keywords of interest to the labeller, notably through leveraging homophilic relationships among hashtags of a same topic (assuming related hashtags are connected). The Latent Dirichlet Allocation (LDA) \cite{blei2003latent} is a traditional and proficient solution to extract topics from a corpus of text documents. Although some works such as \cite{henderson2009applying, xing2016hashtag} have attempted to apply the method to topic discovery on graphs by transforming nodes into documents, and their edges into words, the performance of the resulting algorithm remains low. We found that LDA for graph did not yield meaningful topic distributions when applied to the tweet-hashtag network. Perhaps since each tweet is linked to at most a few hashtags, the small number of resulting words disqualifies LDA as a valid way to model topics on networks with a low median degree. It has also been shown that the topic of keyword expansion is tightly linked to studies on topic discovery \cite{DBLP:journals/corr/abs-2104-09765, tsai2014financial}. However, these studies use methods focusing on large text corpora, which hardly transfer to short text documents such as tweets.

\subsection{Active Learning}
Twitter conversations evolve quickly as events unfold. Active learning provides researchers with tools to dynamically improve the performance of their algorithms over time. \cite{tran2017combination} has leveraged methods to filter relevant Twitter data using active learning in the context of building datasets for Named Entity Recognition tasks. The method was found to greatly help the labeler in their activity of finding related entities using recommendations from a Natural Language Processing algorithm. \cite{rabbany2018active} proposes an active learning method to help investigators gather information against organized criminals by exploring related network information. The authors present a weighted priority scheme allowing to discriminate between different node features, and dynamically re-weight the score of nodes based on incoming human labels. This method appears to be scalable and performs well on real-world datasets. However, it does not directly apply to the keyword evaluation setting as hashtag networks only contain a single feature type, thus discarding the leverage gained through the ranking of multiple features. \cite{bozarth2022keyword} benchmarks the performance of multiple text-based methods \cite{king2017computer, bojanowski2017enriching} to expand a set of seed keywords for the purpose of data collection. They present a high level active keyword expansion setup and show that Word2vec performs better than its baselines. For this reason, we choose to evaluate our graph-based method, \method, against the text-based Word2vec \cite{bojanowski2017enriching}, and Tf-idf methods.

\begin{table*}[t]
\centering
\caption{Summary of the tweet datasets employed in this study. Each row describes a feature of the considered dataset (presented in the columns), or of its retrieved sample. We first describe their full tweet count available for the month of April 2020 in  and then describe the sizes of the small sub-sample leveraged in Section \ref{background}, as well as the full-scale dataset leveraged in Section \ref{sec:exp}. Note that we specify the percentage of the sample datasets that we were able to retrieve next to the tweet count. We also present the number of keywords employed by the authors of the dataset to collect their data.}
\begin{tabular}{ |c|c|c|c| } 
\hline
Dataset Feature & Chen et al. \cite{Chen_2020} & Banda et al. \cite{banda2020largescale} & Qazi et al. \cite{qazi2020geocov19} \\
\hline \hline 
Full Dataset Size & 31,011,964 & 111,406,537 & 271,894,318 \\ \hline
Dataset Keywords Count & 62 & 14 & 716 \\ \hline 
Small Sample Size & 448,100 & 1,543,887 & 3,620,200 \\ \hline 
Small Sample Retrieved Size & 323,363 (72\%) & 1,151,641 (75\%) & 2,649,835 (73\%) \\ \hline 
Full Sample Retrieved Size & 20,800,000 (67\%) & N/A & N/A \\
\hline
\end{tabular}
\label{table:report-full}
\end{table*}

\section{Background}
\label{background}
We first give an overview of past works leveraging tweet datasets to understand the challenges related to collecting Twitter data. We then explore recent open-source datasets to demonstrate the importance of keyword selection when conducting data collection on Twitter.

\subsection{Overview of the Twitter API}
As explained by \cite{ghosh2013sampling, jurgens2016tutorial}, Twitter operates Streaming and Search API services, allowing anyone to collect a free 1\%, live or historical, sample of tweets using queries notably composed of keywords. The recent release of the Twitter API v2 has given academics the capacity to search up to 10 million historical tweets per month without time constraints (as opposed to the previous standard API tier, which only allowed recovery of a small sample of data from the past 7 days). Note that in both cases, users of the Twitter API can leverage keyword queries to retrieve data of interest to their topic of study. These queries are limited either in size (maximum 400 keywords) or character length (256 to 1024 depending on the API tier).

\subsection{Challenges Related to the Twitter API}
The aforementioned API limitations are the first roadblock encountered by researchers while constituting Twitter datasets. A second challenge arises upon sharing their collected data in order to allow for reproducibility or collaboration. Twitter forbids API users from sharing the raw content of the tweet, and only legally allows them to share the tweet's unique identifier. A tweet may be deleted at any time by its author, thus preventing any third party from retrieving it in the future. Not only this may cause trouble upon trying to retrieve data from a published corpus of tweet identifiers (such as for an open-source dataset), but in case of troubles during the initial data collection by researchers, it is hard to make up for the lost data without introducing uncertainty in the corpus of collected tweets by adding more queries. Over time, as tweets are removed by their authors (either by manual deletion or because the author's accounts were deleted), the available tweets included in a dataset diminishes. The efforts required to retrieve a representative sample of the Twitter data employed in academic studies are high. When considering large-scale datasets with millions of tweets, the standard API limits may be quickly reached, slowing the experiments and further reducing the likelihood of accurate reproduction of previous work.

\subsection{Importance of Keyword Selection}
COVID-19 has been a key topic on Twitter in 2020, being the subject of billions of tweets year long. Twitter hosts diverse crowds of agendas and opinions yielding the concurrent spread of public health campaigns and vaccine misinformation. Using the tools provided by Twitter, researchers have assembled large pandemic discussion datasets, gathering billions of tweets \cite{DBLP:journals/corr/abs-2005-03082, Shuja2020.05.19.20107532}. Here, we first present the employed datasets and briefly explain the sampling process. We then compare the datasets to illustrate that in spite of their sizes and a shared topic, their diverging set of keywords leads to non-overlapping sets of tweets.

\subsubsection{COVID-19 Open Source Datasets}
We select three datasets \cite{banda2020largescale, Chen_2020, qazi2020geocov19} to illustrate the role of filtering keywords in the shape of the collected data. Referring to \cite{DBLP:journals/corr/abs-2005-03082} for a comprehensive overview of the available works, we intentionally picked three datasets claiming to use the streaming feature of the standard Twitter API, and reporting different query keyword sets (see table \ref{table:report-full}). Note that further exploration of \cite{Chen_2020} will be conducted in Section \ref{sec:exp}.
We restrain our study of these datasets to the month of April 2020, as the three of them overlap at this time. Moreover, due to their large size, we cannot fully hydrate these datasets. Therefore, we sub-sample our study time frame into 10 random hour slots, located on 10 different days. Note that to ensure that this down-sampling of the dataset study period does not affect the quality of our study, we verify that overall relations between the datasets are conserved when working on our sample.
We report the sizes of the full and sampled datasets in table \ref{table:report-full}. We additionally convert "em" dashes to "en" dashes for consistency. The number of keywords used to collect the data is shown in the third column of table \ref{table:report-full}.
We hydrate the tweet ids from these three sources for the month of April 2020. We were able to retrieve on average 73\% of the tweets from each dataset as some went missing since their original collection by the authors (they may have been deleted either by users or Twitter themselves). 

\subsubsection{Keyword Selection Hypothesis}
\begin{figure}[h]
    \centering
    \includegraphics[width=\columnwidth]{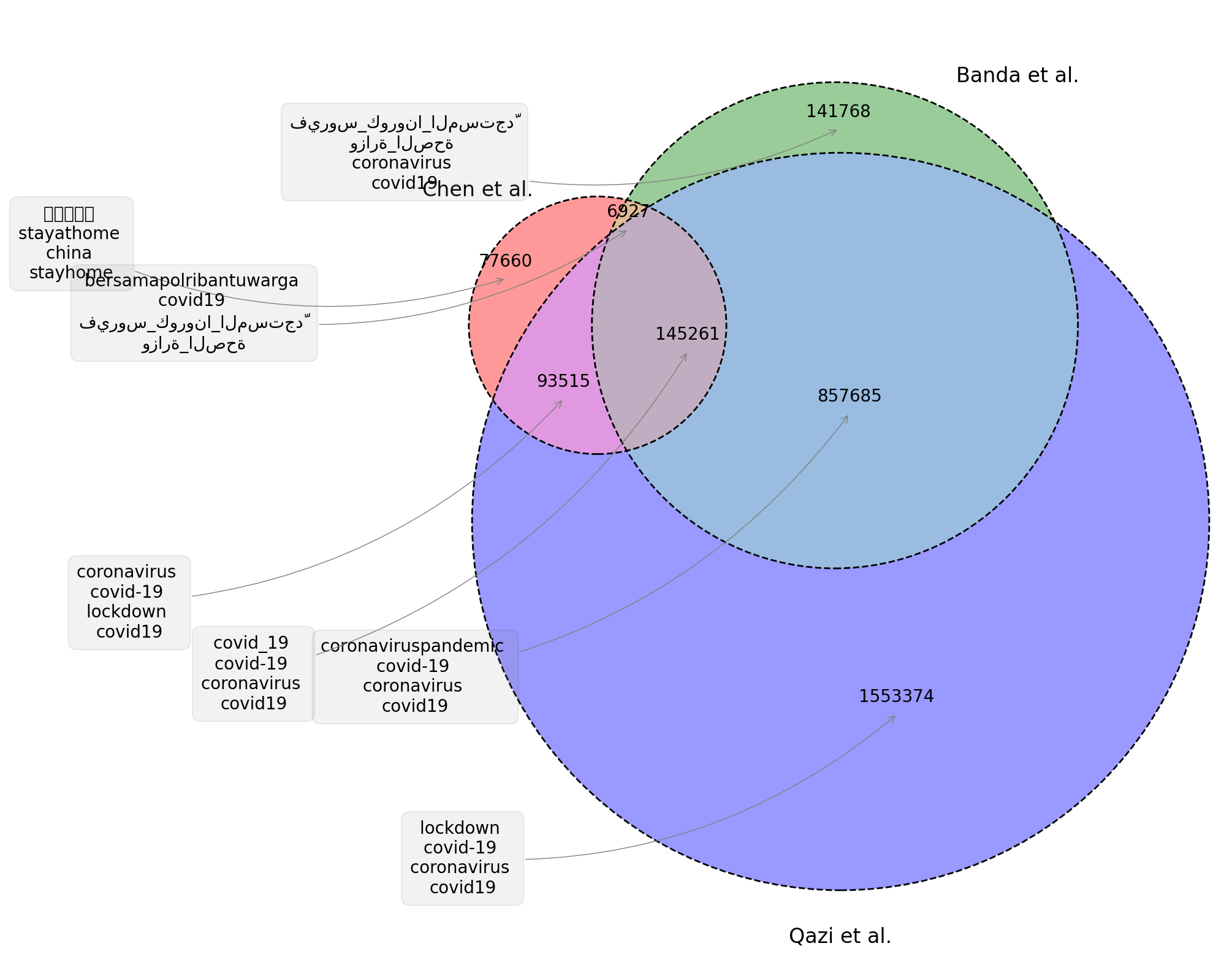}
    \caption{Venn diagram representing tweet overlap across datasets presented in \cite{Chen_2020, qazi2020geocov19, banda2020largescale}. On each set is displayed the number of tweets included in the set, as well as the top hashtags retrieved from the tweets of the set. We first observe that there is no perfect overlap between the three datasets over the sampled periods of time. Secondly, a significant portion of \cite{Chen_2020} is not included in \cite{qazi2020geocov19}, although this dataset employs more keywords, and is much larger by the number of tweets. The most frequent keywords in the excluded portion of \cite{Chen_2020} reveals hashtags related to \textit{stay at home} orders and \textit{lockdown}. With different sets of keywords employed to collect their data, we can conclude that the yielded sets of tweets are impacted by the selected seed keywords.}    \label{fig:datasets_venn}
\end{figure}

Only these 4 hashtags are present among each of the keyword sets employed to collect the aforementioned datasets: \#covid-19, \#coronavirus, \#wuhan, and \#covid19. Thus, there is an almost complete divergence between the keyword sets employed by the authors to retrieve their data. We plot in figure \ref{fig:datasets_venn} the overlap among the hydrated tweets, as described in table \ref{table:report-full}. We observe that above 90\% of the tweets retrieved from \cite{banda2020largescale, Chen_2020} are contained by our sample \cite{qazi2020geocov19}. Conversely, \cite{qazi2020geocov19} only captures 74\% of the tweets sampled from \cite{Chen_2020}. The most popular keywords pertaining to non-overlap regions for each dataset show focus on different issues: one seems to hold a discussion closer to public health campaigns with \textit{lockdown} specific keywords such as \textit{stayathome}, while \cite{banda2020largescale} seems to reach a non-English speaking portion of Twitter with some Arabic top keywords. \cite{qazi2020geocov19} seem to have gathered many more tweets citing the topic of COVID-19 by using over 700 mainstream keywords. As we can observe in figure \ref{fig:datasets_venn}, \cite{Chen_2020} have included public health relevant keywords in their filtering set, and their datasets show differences relative to comparable works. We conclude that the filtering keywords employed by researchers bias the collected data in a non-negligible way. Moreover, a larger volume is not necessarily a solution to these discrepancies. Although \cite{qazi2020geocov19} has retrieved a greater volume of tweets than \cite{Chen_2020}, the lack of representation of public health discussion in their data may weaken downstream analysis.

\section{Method}
\label{sec:meth}
We exploit the user-hashtag graph extracted from a corpus of tweets to help researchers find new, topic-relevant keywords from large Twitter datasets. Tweet-hashtag and user-hashtag graphs can both be extracted from tweets. We focus on the later as it is denser and includes the former, which helps conduct our experiments when few data points are available. We also have referred to both keywords, and hashtags in previous sections. Both terms are used interchangeably in practice, and we will use \textit{hashtags} to describe keywords extracted from tweets, and \textit{keywords} to describe the byproduct of the selection process. We base our method on the following conjectures about keyword-based queries using the search and stream APIs on Twitter:
\begin{compactitem}
    \item User-hashtag networks hold homophily as a feature, thus topic-related hashtags are likely closer to each other than non-topic-related hashtags.
    \item Any topic-relevant keyword is thus more likely to be connected to other topic-relevant keywords, than not.
    \item Extending queries with additional keywords may only increase the volume of yielded tweets.
\end{compactitem}
We outline again that in practice, these conjectures hold based on the quality of the keywords first selected by a researcher to collect their initial tweet dataset. Namely, a keyword set yielding a high volume of tweets with high relevance is more likely to contain new relevant hashtags, than a small set of tweets with low topic relevance.

\subsection{Task}
\label{task}
We define a set of unique tweets, \(T = \{t_1,t_2 \dots t_n\}\), a set of unique users, \(U = \{u_1,u_2 \dots u_k\}\), as well as a set of unique hashtags, \(H = \{H_1,H_2 \dots H_m\}\), with \(H\) representing the adjacency matrix between users contained in \(U\) and hashtag \(j\), such that: \[H_j \in \mathbb{N}^n\ \forall\ j \in [1 \dots m]\]
as could be retrieved from a given tweet (identified by the tweet id, the user id, and mentioning some hashtag).
Given these three sets, we define a \textbf{user-hashtag graph}, with \textit{vertex set}:
\[V = \underbrace{ \{u_1,u_2 \dots u_n\} }_{V_U: user nodes} \cup \underbrace{ \{v_1,v_2 \dots v_m\} }_{V_H: hashtag nodes}\]
where \(\{u_1,u_2 \dots u_n\}\) are \(n\) nodes corresponding to the users from \(U\), and \(\{v_1,v_2 \dots v_m\}\) are the nodes corresponding to hashtags from \(H\), both retrieved from tweet set \(T\).
The \textit{edge set} of the user-hashtag graph is defined as:
\[E=\{(u_i,v_j,[H_j]_i) \mid [H_j]_i > 0 \wedge u_i \in V_U \wedge v_j \in V_H\}\]
where \([H_j]_i \in \mathbb{N}\) is the number of occurrences of hashtag of node \(v_j\) in tweets belonging to user node \(u_i\). We also refer to \([H_j]_i\) as \(w\) for convenience in the rest of the paper.
We define the keyword selection task, as finding hashtag nodes belonging to \(V_H\) that are of interest to the studied topic, determined either through human labeller or a static oracle. As all hashtags pertaining to our set \(H\) are not equally relevant to the topic of interest, we define a set of labelled keywords, of which each element can either be positively labelled, i.e. topic-relevant, or negatively labelled, i.e. not topic-relevant. The labelled keyword set, \(L\), is composed of the union of positive, \(L_+ = \{v_j \mid y_j=1\ \forall\ j \in [1 \dots m]\}\), and negative labelled keywords, \(L_- = \{v_j \mid y_j=-1\ \forall\ j \in [1 \dots m]\}\), such that \(L \equiv L_+ \cup L_-\), and \(y_j \in \{-1,1\}\), a keyword label which can be queried from the oracle. Moreover, we have $v_j \in L_+ \iff y_j = 1$, as well as $v_j \in L_- \iff y_j = -1$.
\begin{definition}[Active Keyword Selection]
Given a set of tweets, \(T\), a set of users, \(U\), and their associated hashtags, \(H\), yielding, \(V_T\),  \(V_U\), \(V_H\), \(E\), as well as an initial set of topic-relevant keywords \(H_s = \{v_j \mid y_j=1\}\), and a fixed labelling budget, \(0 < b << m\), find the maximum number of topic-relevant keywords related to hashtag nodes contained in \(H_s\).
\end{definition}

We explore multiple ways to solve this task in the following sections, starting first by detailing the initialization process adopted to select the candidate keywords to be submitted for labeling to the oracle.

\subsection{Initialization}
\label{init}
Given a corpus of tweets, we extract an edge list connecting hashtags and user ids, \(E\). We sub-sample this edge list using the initial set of positively labelled keywords, \(H_s^a \mid H_s^a \subseteq H\), and retrieve all edges linked to this set of keywords
\[E_s^a = \{(u_i,v_j,[H_{s{_j}}^a]_i) \mid v_j \in H_s^a\}\]
We then expand one-hop along the user dimension of the user-hashtag graph to incorporate all hashtags co-occurring with the initial positively labelled keywords through tweets of a same user:
\[E_s^b = \{(u_i,v_j,[H_{s{_j}}^a]_i) \mid v_j \in H_s^a \vee (v_j \notin H_s^a \wedge \exists e \in E_s^a \mid u_i \in e)\}\]
The resulting edges \(E_s^b\), yield the following set of keywords \[H_s^b = \{v_j \mid (u_i,v_j,[H_{s{_j}}^a]_i) \in E_s^b\}\]
composed of positive keywords from \(H_s^a\), as well as all of their co-occurring hashtags through the sub-sampled edges \(E_s^b\). Note that at this point \(L_+ \equiv H_s^a\).
All keywords \(H_s^b\) are then scored using a given method: \[s\colon (c, N_-(c), N_+(c)) \longmapsto q \mid q \in \mathbb{R}\]
with \(N_-(c) = \{u \mid \exists\ (u,c,w) \in E_s^b \wedge (\exists\ (u,v,w) \in E_s^b \mid v \in L_-)\}\), \(N_+(c) = \{u \mid \exists\ (u,c,w) \in E_s^b \wedge (\exists\ (u,v,w) \in E_s^b \mid v \in L_+)\}\), sets of tweets neighbours both to the candidate keyword \(c\) and labelled keywords. Note that \(s\) is specified in Section \ref{baselines}. All methods presented here follow a breadth-first search, starting by generating a set of candidate keywords: \[C = H_s^b \setminus H_s^a \setminus L \rightarrow C \cap L_- \equiv \emptyset \] where candidate keywords in \(C\) are waiting to be labelled. All candidate keywords are then added to a priority queue, so \(Q\), sorted by their previously determined score.
Note that this process is meant to allow easy experimentation on large datasets encapsulating many sub-topics while reproducing a super-set environment, similar to the sampling environment experienced while collecting data. As researchers come up with an initial set of keywords, the first retrieved user-hashtag network only contains queried and one-hop co-occurring hashtags.

\subsection{Baselines}
\label{baselines}
Each presented baseline, as well our Active Keyword Selection method relies on expanding through the neighborhoods of positively labeled keywords belonging to \(L_+\). The presented baselines generate a score employed to rank a candidate keyword \(c\) in the priority queue \(Q\). For text-based baselines, we retrieve the set of tweet ids yielded by the user-hashtag graph, \(U\), and compile their texts. We do not include texts from retweets for text-based methods.

\subsubsection{Tf-idf}
\label{tfidf}
The term frequency-inverse document frequency measures the importance of a word to a document within a corpora. In our experiment, given a set of tweet $T = \{t_1,t_2,\dots,t_n\}$, which can each be processed into $K_{n,m}$ keywords, we have: 
\[tf = \frac{f_{k,t}}{\sum_{i}|K_{t,i}|}\]
\[idf = \log \frac{|T|}{|\{t \in T | k \in t\}|}\]
\[tfidf(k, t, T) = tf(k,t)*idf(k,T)\]

In order to retrieve a set of candidate keywords with tf-idf, we first give the method a set of pre-processed tweet texts. We sort the keywords by score and retrieve the first $b$ following the budget setting defined in the experiment.

\subsubsection{Word2vec}
\label{word2vec}
The Word2vec model learns a vector representation from a corpus of texts using a neural network model. Based on the co-occurrence of the words in a corpus of text, the word2vec specify the semantic similarity of the words.
Then, it generates the word representations in such a way that there is a high cosine similarity between the representations of the words with similar meaning \cite{mikolov2013efficient}.
In our experiments, we use the package presented by \cite{bojanowski2017enriching} to compute the word embedding from our tweet text corpora. Similar to the Tf-idf method, on each iteration of our experiments we pre-process the selected tweets (sampled using the current seed keywords, $L_+$), and train the model on the resulting documents. We then search for keywords most similar to the the current seed keywords using a cosine similarity metric. We select the $b$ most similar keywords to our seed set of keywords, $L_+$.

\subsubsection{Random Walk}
\label{rw}
Following the previously introduced initialization process, each keyword belonging to \(C\) is attributed a pseudo-random score, defined as follows:
\[s\colon c \longmapsto q \mid q \in [0 \dots 1]\]
The exploration of the positively labeled keywords' neighborhoods is thus done randomly, as the resulting order of the priority queue is pseudo-random. Although this method allows exploring the neighborhoods of low-frequency keywords, it does not leverage any information about positive or negative labeled keywords, or even the topology of the candidate keywords' neighborhood (such as its degree). We explore the latter property in the next baseline.

\subsubsection{Degree Centrality}
\label{degreecentrality}
In some cases, it would be intuitive to guess that topic-relevant keywords would be among the most occurring hashtags of a topic-centered set of tweets. The degree centrality is measured through the degree of the candidate keyword on a unipartite representation of the aforementioned user-hashtag graph. Given a candidate keyword \(c\), we define \[E_c = \{(c, v) \mid (u, c) \in E_s^b \wedge (u, v) \in E_s^b,\ \exists\ u \in V_U, v \in H_s^b \}\]
We can thus define the degree centrality function using the \textit{hashtag-hashtag graph} represented by \(E_c\):
\[s\colon (c, E_c) \longmapsto |E_c|\]
where the cardinality of \(E_c\) is equal to its degree the unipartite graph yielded by the candidate keyword edges. Although this method accounts for the topology of the candidate keyword's neighborhood, it does not leverage potential feedback from either sets of labeled keywords.

\subsection{\method}
Our method follows an iterative labeling process with an initialization as described in Section \ref{init}, where the neighbors of our initial set of positively labeled keywords \(L_+\) are scored and pushed to a priority queue, \(Q\). In this Section, we first specify our novel keyword scoring method and then describe the candidate keyword feedback collection and candidate queue update. We present the general flow in algorithm \ref{alg:aks_algorithm}.

\subsubsection{Weighing Keyword Topic Relevance}
\label{aks}
As introduced in Section \ref{sec:meth}, homophily is assumed to be a key component of the relations between hashtags related to a unique topic. Therefore, we strive to leverage connections of candidate keywords to labeled keywords to discriminate candidates connected to negative keywords, and favor candidates tightly coupled to positive keywords. We define our priority candidate scoring function as
\[s\colon (c, N_-(c), N_+(c)) \longmapsto q \mid q \in \mathbb{R}\]
where \(N_-(c)\), and \(N_+(c)\) represent the sets of tweets neighbors to both the labeled nodes and the candidate keyword. We further specify our function as
\[s(c) = \frac{|N_+(c)|}{|L_+|}-\frac{|N_-(c)|}{|L_-|}\]
where the two terms of the score, evaluate the ratio between the number of tweets linked to both labeled keywords (positive and negative) and the candidate keywords, averaged by the size of the respective labeled keyword set.

\subsubsection{Collecting Feedback on Candidate Keywords}
The iterative process conducted to obtain labels from the oracle on candidate keywords is depicted in algorithm \ref{alg:aks_algorithm}. If the candidate keyword is contained in the oracle set, i.e. topic-relevant, it is added to the set of positively labeled keywords. The hashtag neighbors of the newly labeled candidate keyword are extracted, similarly to the process described in \ref{degreecentrality}, and added to the queue of candidate keywords awaiting labeling. Otherwise, the keyword is added to the set of negatively labeled keywords, i.e. not topic-relevant, and subsequently used as negative feedback for neighboring nodes.

\subsubsection{Computational Efficiency}
\label{efficiency}
The complexity of our candidate priority scoring algorithm is linear in terms of the number of candidate keywords available in the user-hashtag graph:
\[O(|V_H|(|L_+|+|L_-|)) \equiv O(|V_H||L|)\]
Note that we can extend this complexity to relate it to the number of labeling rounds, and the employed budget:
\[O(|V_H|(|H_s^a|+b*k))\]
where \(b\) is the labeling budget employed on every selection round, \(|H_s^a|\) is the number of initially labeled keywords, and \(k\) is the number of selection rounds.

\begin{algorithm}[t]
\small
\caption{Active Keyword Selection Algorithm}
\label{alg:aks_algorithm}
\SetAlgoVlined
\LinesNumbered
\KwData{
$b$, $H_s^a$, $H_s^b$, $E_s^b$, $oracle$
}
\KwResult{$L_{-}$, $L_{+}$}
$L_{-} \gets \{\}$\;
$L_{+} \gets  H_s^a$\;
$C \gets H_s^b \setminus L_- \setminus L_+$\;
$Q \gets \{\}$\;
$N \gets \{v \mid (u, c) \in E_s^b \wedge (u, v) \in E_s^b,\ \forall\ c \in C,\ \exists\ u \in V_U, v \in H_s^b\}$\;
\For{$v \in N$}{
    $score \gets s(v)$\;
    $Q.push(score, v)$\;
}
\While{$b > 0$}{
$c \gets Q.pop()$\;
\If{$c \in oracle$}{
$L_{+} \gets L_{+} + \{c\}$\;
$N \gets \{n \mid (u, c) \in E_s^b \wedge (u, n) \in E_s^b, \exists\ u \in V_U, n \in H_s^b\}$\;
\For{$n \in N$}{
    $score \gets s(n)$\;
    $Q.push(score, n)$\;
}
} 
\Else {
$L_{-} \gets L_{-} + \{c\}$\;
}
$b \gets b - 1$\;
}
\end{algorithm}

\subsection{Evaluation}
\label{sec:metric}
Quantifying the impact of our method on a data collection process is not a trivial task, mostly since running experiments in controlled settings must occur with pre-collected data (see Section \ref{background} for details on the challenges of collecting Twitter data). We assume in this Section a controlled setting with access to the full universe of Twitter data (including topic relevant data), described by the previously mentioned set of tweets $T$, and an oracle set of keywords, $O \mid L_- \in O \wedge L_+ \in O$. We first introduce a definition for two common metrics, \textit{recall}, and \textit{precision}. We then introduce two measures of \textit{user} and \textit{tweet coverage}, employed to understand the performance of the keywords selected in terms of retrieved tweet and user volumes.

\subsubsection{Recall}
\label{sec:recall}
Given a set of oracle keywords, describing the entirety of the topic-relevant keywords, and an initial set of positively labeled keywords $H_s^a$, we assess the capacity of a selection process to retrieve all oracle keywords. We define recall, $r$, such that
\[r = \frac{\big|L_+ \setminus H_s^a \cap O \setminus H_s^a \big|}{\big| O \setminus H_s^a \big|},\ r \in [0 \dots 1]\]
We observe that maximal recall is achieved when $O \subseteq L_+$, meaning that all oracle keywords have been acquired by the set of positively labeled keywords.

\subsubsection{Precision}
\label{sec:precision}
Consider the oracle keyword set $O$, and a set of candidate keywords submitted to the oracle \[ C = L \setminus H_s^a \iff C = (L_+ \cup L_-) \setminus H_s^a\]
where $L$ is the super-set of labeled keywords, and $C$ is the set of keywords that were labeled through a keyword selection as previously presented. We define precision, $p$, as
\[p = \frac{\big| C \cap O \big|}{\big| C \big|} \iff p = \frac{\big| L_+ \big|}{\big| C \big|}\]
where $p \in [0 \dots 1]$, and $H_s^b$ refers to the aforementioned super-set of keywords contained. Thus, maximum precision can only be achieved if all candidate keywords $C$ are contained in $O$. In other terms, precision describes the proportion of positively labeled keywords in the set of candidate keywords.

\subsubsection{User and Tweet Coverage}
\label{coverage}
We define two oracle sets of tweets (\(O_T\)) and users (\(O_U\)) by including any tweet, and therefore user, mentioning an oracle hashtag in their posts. Therefore, given edges $E_U$ from the previously defined user-hashtag graph, and edges $E_T$ from the equivalent tweet-hashtag graph, we measure tweet and user coverage yielded by a set of keywords $v$ as follows:
\[ t_c = \frac{\{t | (t, v) \in E_T | v \in L_+ \}}{|O_T|}\]
\[ u_c = \frac{\{u | (u, v) \in E_U | v \in L_+ \}}{|O_U|}\]
where $u$ and $t$ respectively represent users and tweets, as defined in \ref{task}.
\section{Experiments}
\label{sec:exp}
We first introduce the dataset, and topic-specific oracle keywords employed to conduct our experiments. We then compare the performance of \method to the baselines introduced in Section \ref{baselines}.

\subsection{Dataset}
\label{sec:dataset}
As introduced in Section \ref{background}, we examined three large open-source datasets focusing on the topic of COVID-19 to conduct our study. We selected the work of \cite{Chen_2020} to retrieve at full scale, as the data was both available and complete over the month of April 2020, and maintained up until February 2022, allowing us to consider further extensions of this work. 

\subsubsection{Twitter Universe Approximation}
\label{approximation}
As outlined by \cite{coscia2021noise}, the reproducibility of experiments on Twitter data is contingent on the size of the employed dataset. Given a topic of study, the closer a dataset gets to including all available tweets about the topic of interest, the more likely it is to objectively portray its discussion on Twitter. For the sake of our experiments, we chose to focus on the hegemonic topic of COVID-19, which took Twitter by storm in 2020. As very large datasets have been collected on the topic, we were able to recover the most representative subset of COVID-19 Twitter discussions. Twitter has reported that the hashtag \#COVID19 was tweeted more than 400 million times in 2020 (on average 45,662 times per hour), followed closely by \#StayHome which was the third most tweeted hashtag of the year. With a reported volume of 200 billion per year, the expected 1\% free sample obtainable with the standard Streaming API would be near the 2 billion tweets.

\subsubsection{Tweet hydration}
\label{hydration}
We proceeded to fully hydrate (re-collect tweets using their unique identifiers) the tweets released by \cite{Chen_2020} over the month of April 2020 (see Section \ref{background} for comparison to other open-source datasets). To do so, we loaded all of the 31 million tweet ids in a database, each associated with a hydration status to avoid redundancies. We then use the Twitter Python API, \textit{tweepy}, to lookup tweets in bulk using their tweet ids. Note that we were only able to retrieve 67\% of the reported tweets due to user suspensions and tweet deletions caused by tweet authors and Twitter itself, as previously outlined as a challenge of retrieving Twitter data in Section \ref{background}.

\subsubsection{Study of Three COVID-19 Sub-topics}
\label{subtopics}
We chose to focus on the study of three COVID-19 sub-topics, \textit{Mask}, \textit{Lockdown}, and \textit{Vaccine}, to test and demonstrate the performance of \method. Each of these topics encapsulates complex public health, social, and political issues. Although the topic of "Vaccine" could be considered as "niche" at the time of April 2020, these sub-topics quickly became of crucial importance in the aftermath of the initial spread of the virus, making it an even more interesting point of focus for our experiments. Similarly to \cite{yang2021online}, we derived three sets of oracle keywords, deemed super-sets of relevant keywords for the aforementioned subtopics. We respectively identified 112, 68, and 25 keywords for the topics of \textit{Mask}, \textit{Lockdown}, and \textit{Vaccine}. In terms of popularity, \textit{Lockdown} clearly shows the strongest appeal with more discussions, and keywords with a higher degree (as shown in figure \ref{fig:datasets_venn}). The topic of \textit{Mask} was also found to be popular, as similarly to the real-world implementation of radical measures concerning mandatory quarantine and stay-at-home orders, Personal Protective Equipment (PPE) and masks were a hot-topic at the inception of the pandemic in April 2020. Finally, it was noticeably hard to identify discussions about the topic of \textit{Vaccine} at this time. These topic-specific oracle keywords identified are released alongside the software employed to run these experiments.

\subsection{Performance against Baselines}
The underlying goal of our experimental setting is to simulate a data collection process by running keyword selection over a whole month of data. Every day, all tweets are aggregated and a sample \textit{user-hashtag graph} is generated, following the instructions detailed in Section \ref{sec:meth} to retrieve both \textit{user} and \textit{hashtag} vertex sets $V_U$ and $V_H$, as well as an edge set $E$. We compare \method to its baselines by running active keyword selection processes over 30 days, using three labeling budgets, $\{3, 10, 30\}$.

\subsubsection{Initial Labelled Keyword Selection}
Across all experiment iterations, the oracle keyword sets referred to in Section \ref{subtopics} are maintained constant. In practice, the oracle keyword sets for the topics of \textit{Lockdown}, and \textit{Mask} are sampled to keep only keywords appearing at least once every day throughout the month of data retrieved. We do not sample the oracle keywords for the topic of \textit{Vaccine} as they are already a few keywords and tweets related to the topic.  
Once the oracle keyword set is established, it is sub-sampled to extract the 10 highest degree keywords on the first day of the month, forming an initial set of positively labeled keywords, referred to as $H_s^a$ in Section \ref{init}. Note that this process is coherent with our assumption stated in Section \ref{sec:meth} that topic-relevant keywords with the highest degree are the most likely to be found by researchers upon initiating a data collection process.

\begin{figure}[h]
    \centering
    \includegraphics[width=\columnwidth]{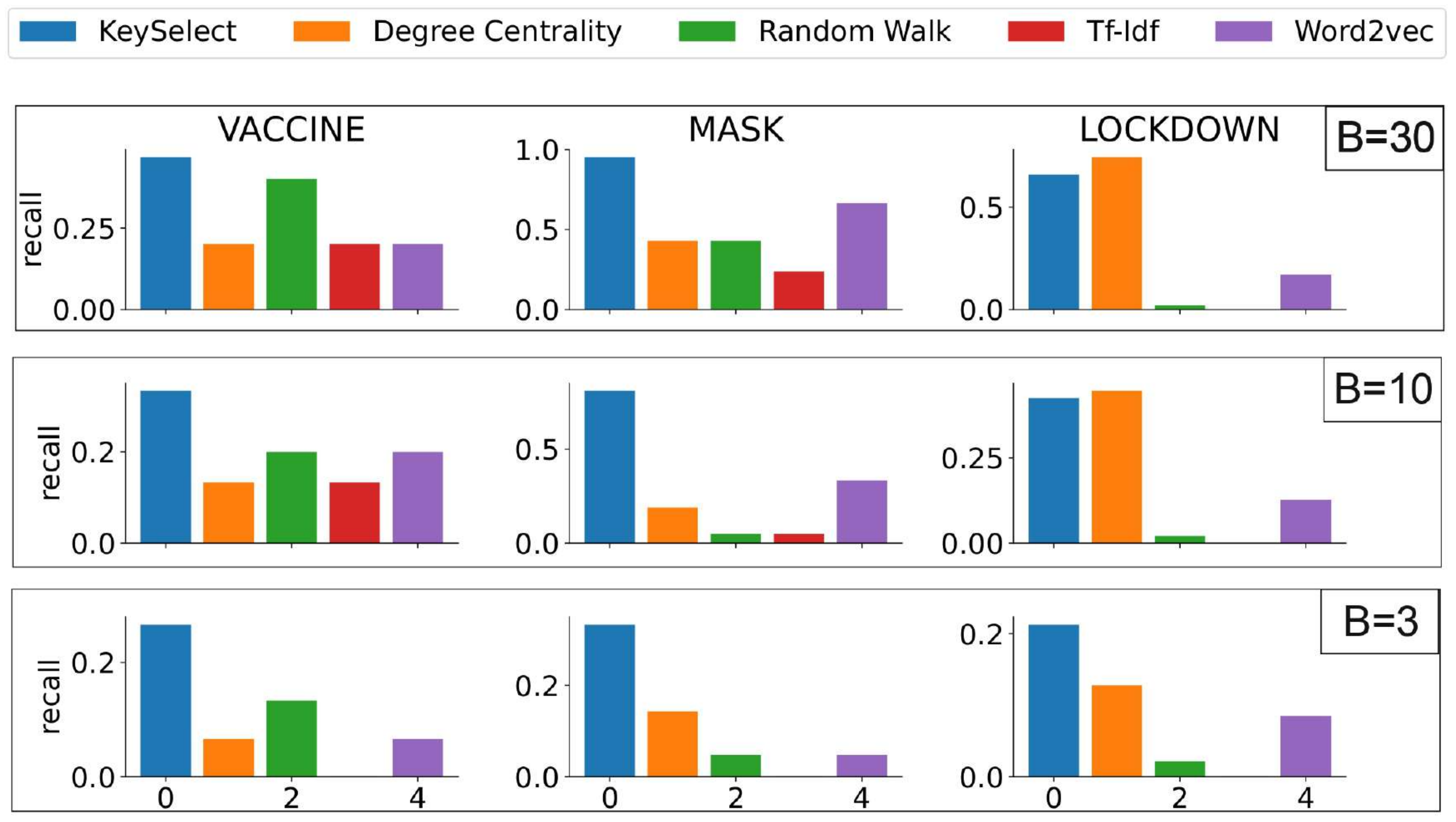}
    \caption{Final recall on keyword selection computed over the month of April 2020 across the three topics of \textit{Vaccine}, \textit{Mask}, and \textit{Lockdown}. Each plot presents the recalls for \method and the baselines presented in \ref{baselines}. A higher recall value indicates that more positive keywords belonging to the oracle set were retrieved.}
    \label{fig:recall}
\end{figure}

\begin{figure}[h]
    \centering
    \includegraphics[width=\columnwidth]{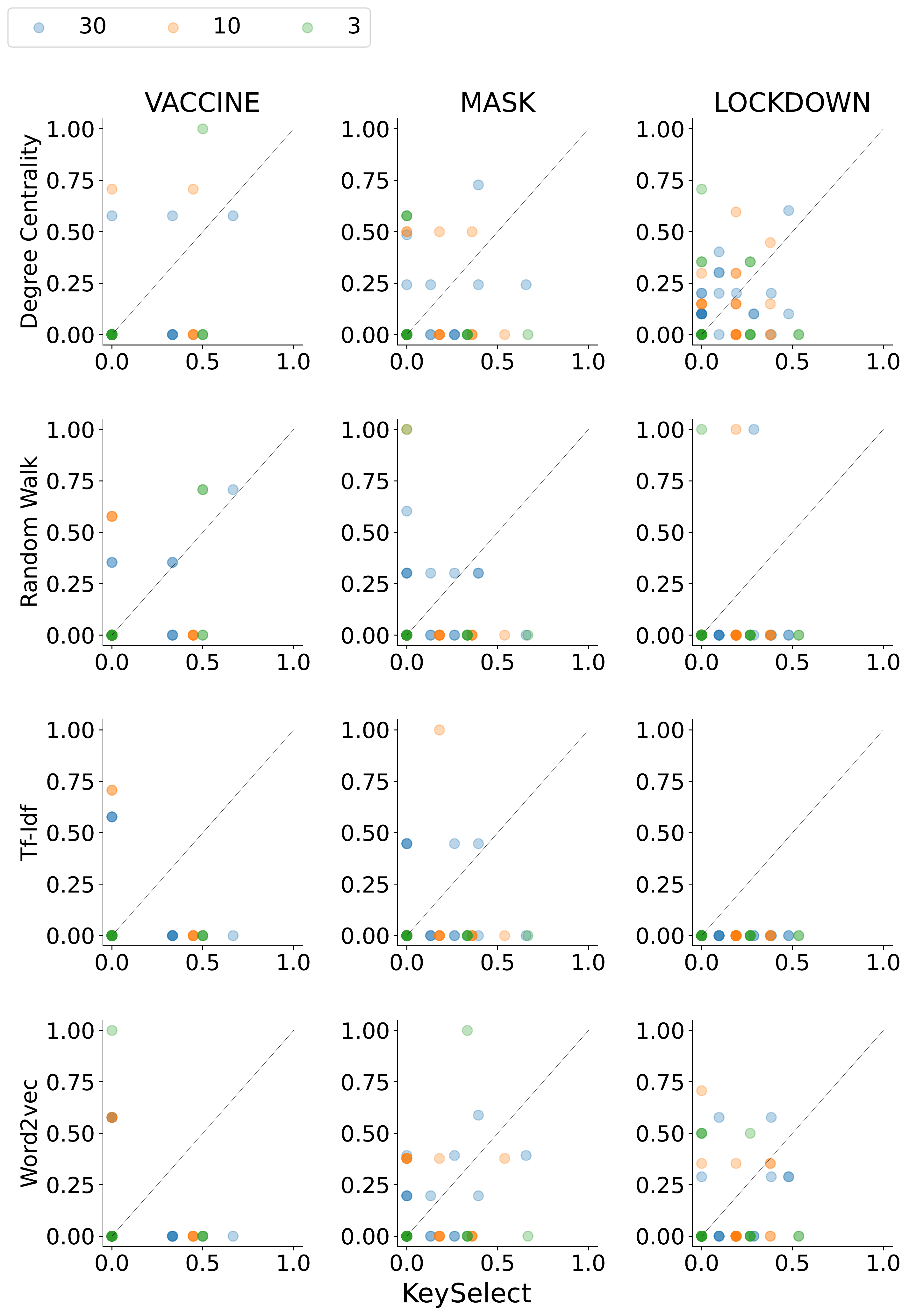}
    \caption{Comparison of normalized precision aggregated over the month of April 2020. The horizontal axis is the precision of \method, and the vertical axis is the precision of a competing baseline. Each point represents the ratio between our method's precision and a given baseline. If the ratio is below 1 (and thus, below the pictured line), \method performs better than the baseline. Each color is associated to a labeling budget. Note that although each budget yields 30 points (for each day), few are visible due to overlap. A higher opacity indicates overlap.}
    \label{fig:precision}
\end{figure}

\subsubsection{Recall and Precision}
We compute the recall as specified in Section \ref{sec:recall}, after completing the active keyword selection process for every configuration of the method, topic, budget, and day of the month. As the recall measures the portion of the oracle keyword set, $O$, retrieved at the end of a selection round, the recall for a given configuration of the method, budget, and topic is monotone increasing over time. We show our results for the final day of the experiment period in figure \ref{fig:recall}. We observe that \method performs better than its baselines in all settings, except when considering the topic of \textit{Lockdown} with budgets of $\{10, 30\}$. Similarly, the precision measure is computed as specified in Section \ref{sec:precision}. The pairwise comparison of normalized precision between \method and its baselines is represented in Figure \ref{fig:precision}. We observe that similarly to the recall measure, our method has greater precision than its baselines, except against Degree Centrality on the topic of \textit{Lockdown}. Since the oracle set is static, there are only few possible precision values for every experiment configuration.

\begin{table*}
\centering
\caption{Tweet coverage scores computed by topic and budget. We compute the tweet coverage score as presented in Section \ref{coverage} on the last iteration of the Active Keyword Selection process. Each column corresponds to a combination of settings, with topic (\textit{Vaccine} is abbreviated to \textit{Vacc.} for formatting) and budget values.}
\begin{tabular}{ |c|c|c|c|c|c|c|c|c|c| } 
\hline
Method & \textit{Vacc.} (3) & \textit{Vacc.} (10) & \textit{Vacc.} (30) & \textit{Mask} (3) & \textit{Mask} (10) & \textit{Mask} (30) & \textit{Lockdown} (3) & \textit{Lockdown} (10) & \textit{Lockdown} (30)  \\
\hline \hline 
Degree Centrality & 0.86 & 0.86 & 0.89 & 0.79 & 0.81 & 0.84 & \textbf{0.88} & \textbf{0.91} & 0.94 \\ \hline
\method & \textbf{0.89} & \textbf{0.90} & \textbf{0.91} & \textbf{0.82} & \textbf{0.87} & \textbf{0.91} & 0.87 & 0.89 & \textbf{0.95} \\ \hline 
Random Walk & 0.86 & 0.84 & 0.90 & 0.71 & 0.71 & 0.79 & 0.86 & 0.86 & 0.86 \\ \hline 
Tf-Idf \cite{bozarth2022keyword} & 0.83 & 0.88 & 0.89 & 0.71 & 0.71 & 0.73 & 0.86 & 0.86 & 0.86 \\ \hline 
Word2Vec \cite{bojanowski2017enriching} & 0.84 & 0.85 & 0.85 & 0.71 & 0.76 & 0.83 & 0.87 & 0.87 & 0.89 \\
\hline
\end{tabular}
\label{table:tweet-coverage}
\end{table*}

\begin{table*}
\centering
\caption{User coverage scores computed by topic and budget. We compute the user coverage score as presented in Section \ref{coverage} on the last iteration of the Active Keyword Selection process. Each column corresponds to a combination of settings, with topic and budget values.}
\begin{tabular}{ |c|c|c|c|c|c|c|c|c|c| } 
\hline
Method & \textit{Vacc.} (3) & \textit{Vacc.} (10) & \textit{Vacc.} (30) & \textit{Mask} (3) & \textit{Mask} (10) & \textit{Mask} (30) & \textit{Lockdown} (3) & \textit{Lockdown} (10) & \textit{Lockdown} (30)  \\
\hline \hline 
Degree Centrality & 0.86 & 0.86 & 0.89 & 0.80 & 0.81 & 0.85 & \textbf{0.88} & \textbf{0.91} & \textbf{0.95} \\ \hline
\method & \textbf{0.90} & \textbf{0.90} & \textbf{0.91} & \textbf{0.83} & \textbf{0.87} & \textbf{0.91} & 0.87 & 0.90 & 0.94 \\ \hline 
Random Walk & 0.86 & 0.84 & 0.90 & 0.70 & 0.70 & 0.79 & 0.86 & 0.86 & 0.87 \\ \hline 
Tf-Idf \cite{bozarth2022keyword} & 0.82 & 0.88 & 0.89 & 0.70 & 0.70 & 0.72 & 0.86 & 0.86 & 0.86 \\ \hline 
Word2Vec \cite{bojanowski2017enriching} & 0.84 & 0.85 & 0.85 & 0.70 & 0.76 & 0.83 & 0.87 & 0.87 & 0.90 \\
\hline
\end{tabular}
\label{table:user-coverage}
\end{table*}

\subsubsection{User and Tweet Coverage}
The coverage measures are computed as specified in Section \ref{coverage}. As presented in tables \ref{table:tweet-coverage} and \ref{table:user-coverage}, we find that our method significantly outperforms the text-based baselines in the volume of retrieved tweets and users from the oracle sets, and consistently matches the performance of the best graph-based baselines.

\subsection{Discussion - practitioner's guide}
The degree distribution of hashtags yielded by the graph with edges $E$ described in Section \ref{task} follows a power law. Thus, a fat tail contains many of the keywords of the dataset, with few keywords capturing large occurrence frequencies. As seen in Section \ref{background}, \textit{Lockdown} is a dominant topic within the dataset \cite{Chen_2020}, thus implying that its topic-relevant hashtags, and therefore, our oracle keywords, are contained among these most occurring keywords, and less frequent among hashtags contained in the fat-tail portion of the distribution. This may explain why the Degree Centrality performs better than \method at retrieving topic-relevant keywords contained in the oracle on more popular topics, such as \textit{Lockdown}. Note that this comes as a limitation of our experimental setting, where the resulting hashtag occurrence frequency distribution on less prevalent sub-topics does not necessarily fit the distribution presumably yielded by the Twitter API.
\section{Conclusions}
\label{sec:concl}
We presented \method, which addresses the keyword selection problem using active learning to help researchers find topic-relevant keywords to conduct data collection on Twitter.
The advantages of \method are:
\begin{itemize}
    \item{ {\bf \effective}: it yields higher recall, as well as user and tweet coverages than baselines thus allowing to retrieve datasets of higher quality.}
    \item{ {\bf \automatic}: its only parameter is the labeling budget. We suggest a reasonable default of 30, but \method has shown superior performance compared with baselines irrespective of the budget.}
\end{itemize}
We presented experiments on a dataset of 21 million tweets, where \method demonstrated its capacity to retrieve topic-relevant keywords for topics of diverse popularity, while notably outperforming text-based baselines presented by \cite{bozarth2022keyword}.
\method comes as a solution to the problem of Active Keyword Selection, by allowing labelers to easily access topic-relevant keywords. In practice, an implementation of this method may save a researcher time by automating the keyword suggestion process, avoiding manual exhaustive search either of Twitter itself or through unorganized pre-collected data.

{\bf Usage and Reproducibility:} We have already open-sourced the code for our method and experiments at \url{https://github.com/sachalevy/active-keyword-selection}, as well as a web interface to use \method.

\bibliographystyle{IEEEtran}

\bibliography{main}

\begin{thebibliography}{10}
\providecommand{\url}[1]{#1}
\csname url@samestyle\endcsname
\providecommand{\newblock}{\relax}
\providecommand{\bibinfo}[2]{#2}
\providecommand{\BIBentrySTDinterwordspacing}{\spaceskip=0pt\relax}
\providecommand{\BIBentryALTinterwordstretchfactor}{4}
\providecommand{\BIBentryALTinterwordspacing}{\spaceskip=\fontdimen2\font plus
\BIBentryALTinterwordstretchfactor\fontdimen3\font minus
  \fontdimen4\font\relax}
\providecommand{\BIBforeignlanguage}[2]{{%
\expandafter\ifx\csname l@#1\endcsname\relax
\typeout{** WARNING: IEEEtran.bst: No hyphenation pattern has been}%
\typeout{** loaded for the language `#1'. Using the pattern for}%
\typeout{** the default language instead.}%
\else
\language=\csname l@#1\endcsname
\fi
#2}}
\providecommand{\BIBdecl}{\relax}
\BIBdecl

\bibitem{kejriwal2018pipeline}
M.~Kejriwal and Y.~Gu, ``A pipeline for post-crisis twitter data acquisition,''
  2018.

\bibitem{banda2020largescale}
J.~M. Banda, R.~Tekumalla, G.~Wang, J.~Yu, T.~Liu, Y.~Ding, K.~Artemova,
  E.~Tutubalina, and G.~Chowell, ``A large-scale covid-19 twitter chatter
  dataset for open scientific research -- an international collaboration,''
  2020.

\bibitem{alqurashi2020large}
S.~Alqurashi, A.~Alhindi, and E.~Alanazi, ``Large arabic twitter dataset on
  covid-19,'' 2020.

\bibitem{Chen_2020}
\BIBentryALTinterwordspacing
E.~Chen, K.~Lerman, and E.~Ferrara, ``Tracking social media discourse about the
  covid-19 pandemic: Development of a public coronavirus twitter data set,''
  \emph{JMIR Public Health and Surveillance}, vol.~6, no.~2, p. e19273, May
  2020. [Online]. Available: \url{http://dx.doi.org/10.2196/19273}
\BIBentrySTDinterwordspacing

\bibitem{chen2020election2020}
E.~Chen, A.~Deb, and E.~Ferrara, ``\#election2020: The first public twitter
  dataset on the 2020 us presidential election,'' 2020.

\bibitem{achrekar2011predicting}
H.~Achrekar, A.~Gandhe, R.~Lazarus, S.-H. Yu, and B.~Liu, ``Predicting flu
  trends using twitter data,'' in \emph{2011 IEEE conference on computer
  communications workshops (INFOCOM WKSHPS)}.\hskip 1em plus 0.5em minus
  0.4em\relax IEEE, 2011, pp. 702--707.

\bibitem{jurgens2016tutorial}
P.~J{\"u}rgens and A.~Jungherr, ``A tutorial for using twitter data in the
  social sciences: Data collection, preparation, and analysis,''
  \emph{Preparation, and Analysis (January 5, 2016)}, 2016.

\bibitem{pfeffer2018tampering}
J.~Pfeffer, K.~Mayer, and F.~Morstatter, ``Tampering with twitter’s sample
  api,'' \emph{EPJ Data Science}, vol.~7, no.~1, p.~50, 2018.

\bibitem{morstatter2016tamper}
\BIBentryALTinterwordspacing
F.~Morstatter, H.~Dani, J.~Sampson, and H.~Liu, ``Can one tamper with the
  sample api? toward neutralizing bias from spam and bot content,'' in
  \emph{Proceedings of the 25th International Conference Companion on World
  Wide Web}, ser. WWW '16 Companion.\hskip 1em plus 0.5em minus 0.4em\relax
  Republic and Canton of Geneva, CHE: International World Wide Web Conferences
  Steering Committee, 2016, p. 81–82. [Online]. Available:
  \url{https://doi.org/10.1145/2872518.2889372}
\BIBentrySTDinterwordspacing

\bibitem{bozarth2022keyword}
L.~Bozarth and C.~Budak, ``Keyword expansion techniques for mining social
  movement data on social media,'' \emph{EPJ Data Science}, vol.~11, no.~1,
  p.~30, 2022.

\bibitem{qazi2020geocov19}
U.~Qazi, M.~Imran, and F.~Ofli, ``Geocov19: A dataset of hundreds of millions
  of multilingual covid-19 tweets with location information,'' 2020.

\bibitem{ghosh2013sampling}
S.~Ghosh, M.~B. Zafar, P.~Bhattacharya, N.~Sharma, N.~Ganguly, and K.~Gummadi,
  ``On sampling the wisdom of crowds: random vs. expert sampling of the twitter
  stream,'' in \emph{Proceedings of the 22nd ACM international conference on
  Information \& Knowledge Management}, 2013, pp. 1739--1744.

\bibitem{Shuja2020.05.19.20107532}
\BIBentryALTinterwordspacing
J.~Shuja, E.~Alanazi, W.~Alasmary, and A.~Alashaikh, ``Covid-19 open source
  data sets: A comprehensive survey,'' \emph{medRxiv}, 2020. [Online].
  Available:
  \url{https://www.medrxiv.org/content/early/2020/07/13/2020.05.19.20107532}
\BIBentrySTDinterwordspacing

\bibitem{yang2021online}
Z.~Yang, A.~Imouza, K.~Pelrine, S.~L{\'e}vy, J.~Liu, G.~Desrosiers-Brisebois,
  J.-F. Godbout, A.~Blais, and R.~Rabbany, ``Online partisan polarization of
  covid-19,'' in \emph{2021 International Conference on Data Mining Workshops
  (ICDMW)}.\hskip 1em plus 0.5em minus 0.4em\relax IEEE, 2021, pp. 893--901.

\bibitem{marujo2015automatic}
L.~Marujo, W.~Ling, I.~Trancoso, C.~Dyer, A.~W. Black, A.~Gershman, D.~M.
  de~Matos, J.~P. Neto, and J.~G. Carbonell, ``Automatic keyword extraction on
  twitter,'' in \emph{Proceedings of the 53rd Annual Meeting of the Association
  for Computational Linguistics and the 7th International Joint Conference on
  Natural Language Processing (Volume 2: Short Papers)}, 2015, pp. 637--643.

\bibitem{king2017computer}
G.~King, P.~Lam, and M.~E. Roberts, ``Computer-assisted keyword and document
  set discovery from unstructured text,'' \emph{American Journal of Political
  Science}, vol.~61, no.~4, pp. 971--988, 2017.

\bibitem{leskovecfaloustos2006}
J.~Leskovec and C.~Faloutsos, ``Sampling from large graphs,'' 2006.

\bibitem{hu2013survey}
P.~Hu and W.~C. Lau, ``A survey and taxonomy of graph sampling,'' 2013.

\bibitem{wang2011understanding}
T.~Wang, Y.~Chen, Z.~Zhang, T.~Xu, L.~Jin, P.~Hui, B.~Deng, and X.~Li,
  ``Understanding graph sampling algorithms for social network analysis,'' in
  \emph{2011 31st international conference on distributed computing systems
  workshops}.\hskip 1em plus 0.5em minus 0.4em\relax IEEE, 2011, pp. 123--128.

\bibitem{stryker2006}
J.~Stryker, R.~Wray, R.~Hornik, and I.~Yanovitzky,
  ``\BIBforeignlanguage{English (US)}{Validation of database search terms for
  content analysis: The case of cancer news coverage},''
  \emph{\BIBforeignlanguage{English (US)}{Journalism and Mass Communication
  Quarterly}}, vol.~83, no.~2, pp. 413--430, Jan. 2006.

\bibitem{Krippendorff2011ComputingKA}
K.~Krippendorff, ``Computing krippendorff's alpha-reliability,'' 2011.

\bibitem{blei2003latent}
D.~M. Blei, A.~Y. Ng, and M.~I. Jordan, ``Latent dirichlet allocation,''
  \emph{Journal of machine Learning research}, vol.~3, no. Jan, pp. 993--1022,
  2003.

\bibitem{henderson2009applying}
K.~Henderson and T.~Eliassi-Rad, ``Applying latent dirichlet allocation to
  group discovery in large graphs,'' in \emph{Proceedings of the 2009 ACM
  symposium on Applied Computing}, 2009, pp. 1456--1461.

\bibitem{xing2016hashtag}
C.~Xing, Y.~Wang, J.~Liu, Y.~Huang, and W.-Y. Ma, ``Hashtag-based sub-event
  discovery using mutually generative lda in twitter,'' in \emph{Proceedings of
  the AAAI Conference on Artificial Intelligence}, vol.~30, no.~1, 2016.

\bibitem{DBLP:journals/corr/abs-2104-09765}
\BIBentryALTinterwordspacing
Y.~Jin, A.~Bhatia, and D.~Wanvarie, ``Seed word selection for weakly-supervised
  text classification with unsupervised error estimation,'' \emph{CoRR}, vol.
  abs/2104.09765, 2021. [Online]. Available:
  \url{https://arxiv.org/abs/2104.09765}
\BIBentrySTDinterwordspacing

\bibitem{tsai2014financial}
M.-F. Tsai and C.-J. Wang, ``Financial keyword expansion via continuous word
  vector representations,'' in \emph{Proceedings of the 2014 Conference on
  Empirical Methods in Natural Language Processing (EMNLP)}, 2014, pp.
  1453--1458.

\bibitem{tran2017combination}
V.~C. Tran, N.~T. Nguyen, H.~Fujita, D.~T. Hoang, and D.~Hwang, ``A combination
  of active learning and self-learning for named entity recognition on twitter
  using conditional random fields,'' \emph{Knowledge-Based Systems}, vol. 132,
  pp. 179--187, 2017.

\bibitem{rabbany2018active}
R.~Rabbany, D.~Bayani, and A.~Dubrawski, ``Active search of connections for
  case building and combating human trafficking,'' in \emph{Proceedings of the
  24th ACM SIGKDD International Conference on Knowledge Discovery \& Data
  Mining}, 2018, pp. 2120--2129.

\bibitem{bojanowski2017enriching}
P.~Bojanowski, E.~Grave, A.~Joulin, and T.~Mikolov, ``Enriching word vectors
  with subword information,'' \emph{Transactions of the association for
  computational linguistics}, vol.~5, pp. 135--146, 2017.

\bibitem{DBLP:journals/corr/abs-2005-03082}
\BIBentryALTinterwordspacing
C.~Ordun, S.~Purushotham, and E.~Raff, ``Exploratory analysis of covid-19
  tweets using topic modeling, umap, and digraphs,'' \emph{CoRR}, vol.
  abs/2005.03082, 2020. [Online]. Available:
  \url{https://arxiv.org/abs/2005.03082}
\BIBentrySTDinterwordspacing

\bibitem{mikolov2013efficient}
T.~Mikolov, K.~Chen, G.~Corrado, and J.~Dean, ``Efficient estimation of word
  representations in vector space,'' \emph{arXiv preprint arXiv:1301.3781},
  2013.

\bibitem{coscia2021noise}
M.~Coscia, ``Noise corrected sampling of online social networks,'' \emph{ACM
  Transactions on Knowledge Discovery from Data (TKDD)}, vol.~15, no.~2, pp.
  1--21, 2021.

\end{thebibliography}

\end{document}